\documentclass[sigconf]{acmart}

\usepackage{enumitem}

\makeatletter
\def\@ACM@checkaffil{
    \if@ACM@instpresent\else
    \ClassWarningNoLine{\@classname}{No institution present for an affiliation}%
    \fi
    \if@ACM@citypresent\else
    \ClassWarningNoLine{\@classname}{No city present for an affiliation}%
    \fi
    \if@ACM@countrypresent\else
        \ClassWarningNoLine{\@classname}{No country present for an affiliation}%
    \fi
}
\makeatother
\AtBeginDocument{%
  \providecommand\BibTeX{{%
    \normalfont B\kern-0.5em{\scshape i\kern-0.25em b}\kern-0.8em\TeX}}}

\setcopyright{acmcopyright}
\copyrightyear{2022}
\acmYear{2023}
\acmDOI{}

\acmConference[]{}{February 27-March 3,
  2023}{Singapore}
\begin{document}

\title{Gender Bias in Fake News: An Analysis}

\author{Navya Sahadevan$^1$ \hspace{0.1in} Deepak P$^2$}
\affiliation{$^1$St. Joseph's College, Devagiri, Kerala, India \\ $^2$Queen's University Belfast, UK \\ navyapulkurni@gmail.com \hspace{0.2in} deepaksp@acm.org}


\begin{abstract}
Data science research into fake news has gathered much momentum in recent years, arguably facilitated by the emergence of large public benchmark datasets. While it has been well-established within media studies that gender bias is an issue that pervades news media, there has been very little exploration into the relationship between gender bias and fake news. In this work, we provide the first empirical analysis of gender bias vis-a-vis fake news, leveraging simple and transparent lexicon-based methods over public benchmark datasets. Our analysis establishes the increased prevalence of gender bias in fake news across three facets viz., abundance, affect, political terms and proximal words. The insights from our analysis provide a strong argument that gender bias needs to be an important consideration in research into fake news.
\end{abstract}





\maketitle

\section{Introduction}

{\it Fake news}, a terminology in use since 1890s\footnote{https://www.merriam-webster.com/words-at-play/the-real-story-of-fake-news}, refers to false or misleading content presented as news. This has been a subject of intense academic and popular interest lately, especially coinciding with the emergence of online media, and professional management of political activity using it~\cite{johnson2016campaigning}. Disinformation or information disorder, as the phenomenon has been recently referred to, has been increasingly recognized as a grave threat to deliberative democracy~\cite{mckay2021disinformation}. While relationship between fake news and politics has rightly been subject to much scrutiny, fake news has been recently observed within the context of the COVID-19 pandemic~\cite{springer2022disinformation} as well. In this paper, we analyze the relationship between fake news and gender, one that has been recognized within social sciences viz., media studies~\cite{almenar2021gender} and politics~\cite{stabile2019sex}, and needs to be explored within quantitative Natural Language Processing (NLP). 

In this paper, for the first time to our best knowledge, we consider differences in how gender identities are portrayed within fake and real news within popular public textual fake news benchmark datasets. In particular, our interest is in understanding the relative abundance of gender references as well as gender-specific trends on prevalence of affect and political terms, as well as and lexical references in general, and how they vary across fake and real news. Our analysis illustrates the nuanced but consistent nature of the relationship between news veracity and gender, and strongly suggests that gender needs to be an important consideration within critical studies and data science research on fake news.

\section{Related Work}\label{sec:relatedwork}

It is interesting to note that what may be regarded as the watershed moment for the fake news phenomenon - 2016 US presidential election~\cite{grinberg2019fake} - was a contest between female and male contenders. Yet, analysis focusing on gender has been very limited. A notable study~\cite{stabile2019sex} touching on the gender aspect during the 2016 election uses quantitative analysis of tweets surrounding the election providing evidence that election-time fake news had supported stereotypes such as women being unfit for leadership positions, hand in hand with villianizing or trivializing women. In contrast to this work centered on a particular election, our study has a broader scope and considers gender bias over popular benchmark datasets. 

Research into gender and fake news located within media studies as well as broadly within social sciences have uncovered interesting insights. A survey study over Spanish respondents~\cite{almenar2021gender} found that the perceptions of fake news varied across genders in terms of the degrees of concern, but remained underpinned by the same problems. An in-depth and comprehensive deliberation of the issues around disinformation within the context of race and gender~\cite{thakur2022facts} - published in 2022 - raises several interesting points. It suggests that disinformation flows from the same patriarchal context as online gender-based violence and that gendered disinformation seeks to reinforce negative views of women. In other words, gender stereotypes and biases have been argued to form another dimension for disinformation. It is to be noted that such gender issues have even been noted in pandemic disinformation~\cite{sessa2020misogyny}. This backdrop of extant social science research into gender and disinformation motivates our empirical study. 

Against the backdrop of such social science literature, our attempt is to take a few early steps in data-driven quantitative studies into analyzing the relationships between gender and textual disinformation using popular public datasets leveraged within data science research on fake news.

\section{Research Questions}

Our research questions on data-driven analysis of gender vis-a-vis fake news are:

\begin{itemize}[leftmargin=*]
\setlength\itemsep{0em}
\item {\bf RQ1:} How do gender groups fare on their relative abundance within fake and real (i.e., non-fake) news? \\
\item {\bf RQ2:} How do emotions, political references and sentiments differ around gender mentions across fake and real news? \\
\item {\bf RQ3:} How do trends on lexical references surrounding references to gender groups differ across fake and real news?
\end{itemize}

\noindent We explore these questions over large benchmark datasets for fake news research. 

\section{Analysis Methodology}


We first start with outlining the basic building blocks we use for our analysis. We intent that our analysis methodology embeds the ability to be be clearly verifiable by humans and be able to trace back the inferences to particular mentions within the textual articles. This makes lexicon-based statistical approaches more suited than ML-based approaches for the task. Thus, our building blocks use simple and transparent lexicon-based analysis strategies. These building blocks, as we will see, would be used in a straightforward way for the analysis to address the separate RQs. 


\noindent{\bf Gender Mentions:} The first key component in our method involves the identification of gender references/mentions within (fake and real) textual news articles. We use a {\it gender reference lexicon} extracted from the NLTK corpus\footnote{https://www.nltk.org/api/nltk.corpus.html} which contains separate lists of mentions of {\it male} and {\it female} genders, which includes gender-correlated names and pronouns. For example, words such as {\it he} and {\it his} are part of the {\it male reference list}, whereas {\it her} and {\it girl} are part of the {\it female reference list}. The separate lists contained thousands of words each. Lexicon word occurrences within news articles are regarded as respective gender mentions. 


\noindent{\bf Emotions, Political References and Sentiments:} Similar to identification of gender mentions, we use the popular word-emotion association lexicon, {\it EmoLex}\footnote{http://www.saifmohammad.com/}, to identify affective terms within textual news articles across eight emotion classes, viz., {\it anger, anticipation, disgust, fear, joy, sadness, surprise} and {\it trust}. Given our intent of analyzing emotions and sentiments against the backdrop of gender references, we identify emotion mentions within a specified width word window - the width being a hyperparameter - on either side of the gender references. As an example, within the text excerpt: {\it 'there was apparent anger in her face'}, the emotion word {\it anger} from the {\it anger} emotion class would be associated with the female gender pronoun {\it her} for window sizes set to $\geq 2$; we did our analysis with window size varying from $5$ to $10$. We use the same methodology for identifying political references, our political lexicon compiled from various sources to include all types of political references including current affairs made publicly available on GitHub\footnote{https://github.com/pulkurni/politcal-lexcicon}.Towards analyzing sentiments, we leveraged a popular Python lexicon-based sentiment library, VADER\footnote{https://pypi.org/project/vaderSentiment/}. We aggregate the positive/negative sentiment-scores provided for mention-proximal words that appear in VADER's sentiment dictionaries, to arrive at cumulative sentiment polarities around each gender mention. 

\noindent{\bf Proximal Lexical References:} In analyzing the nature of lexical references surrounding gender mentions, we use the same window-based approach as earlier, and associate all words within a fixed width word window to the gender reference. 



\noindent{\bf RQ-specific Analyses:} For a news article $N$, let the bag of {\it female} and {\it male} gender mentions be $N_F$ and $N_M$ respectively. The bag of emotion words from emotion class $e$ taken separately around female references in $N$, denoted as $N_F^e$ would be:

\[ N_F^e = \mathop{\cup}_{m \in N_F} \{ m_e | m_e \in L(E=e) \wedge m_e \in W(m) \} \]


where $W(m)$ denotes the adjacency window around the mention $m$, and $L(E=e)$ denotes the lexicon for emotion class $e$; all set notations above be interpreted as bag notations. $N_M^e$, the bag of emotion words around male mentions, is defined similarly. 


The overall emotion density around female mentions in $N$ is measured as:

\[ N_F^{ED} = \frac{\sum_{e \in E} |N_F^e|}{|N_F^W|} \]

For political references, we denote the bag of political words around {\it female} mentions as:

\[ N_F^P = \mathop{\cup}_{m \in N_F} \{ m_p | m_p \in \mathcal{PL} \wedge m_p \in W(m) \} \]

where $\mathcal{PL}$ is the political lexicon employed. $N_M^P$, the bag of political terms around male mentions, is defined similarly. As in the case of emotion densities, the political reference density is computed as:

\[ N_F^{PD} = \frac{|N_F^P|}{|N_F^W|} \]


\noindent For sentiment analysis, a sentiment score is computed for each gender mention $m$:

\begin{equation}
S(m) = \sum_{w \in W(m)} VADER(w)
\end{equation}

where $VADER(w)$ denotes positive or negative sentiment score from VADER ($0.0$ assumed if $w \not\in VADER$). Each mention is then associated with one of three sentiment labels, {\it Positive}, {\it Neutral} and {\it Negative}, based on the sentiment score. All such news article specific measures would be suitably aggregated over subsets of {\it real} and {\it fake} news. The results are then normalised and converted into percentage to address RQ1-3, as we will describe.




\section{Results and Analysis}



\noindent{\bf Datasets and Setup:} In the interest of generality, we conduct our analysis over four popular textual datasets that have been used for fake news research. ISOT\footnote{https://www.uvic.ca/ecs/ece/isot/datasets/fake-news/index.php} comprises 40k+ articles split roughly evenly between {\it real} and {\it fake}. LIAR~\cite{wang2017liar} comprises 12.8k statements labelled for veracity. FakeNewsNet~\cite{shu2020fakenewsnet} (FNN) has 21k news articles collected from across Politifact and Gossipcop. The smallest dataset, BuzzFeed~\cite{potthast-etal-2018-stylometric} comprises a complete sample of news published in Facebook over a week close to the 2016 U.S. election.  There is much variety in cardinality, type, distribution and source of news articles across these three datasets, facilitating a well-rounded analysis across them. Our observed trends were largely invariant with window size variations; the results reported are with window size set to $5$.




\begin{table}[]
      \centering
        \begin{tabular}{|c|c|c|c|}
        \hline
        {\bf Data} & {\bf Ver.} & {\bf F} & {\bf M} \\
        \hline
        ISOT
            & Real & 25 & 75 \\
            \cline{2-4}
            & Fake & 25 & 75 \\
        \hline
        LIAR
            & Real & 24 & 76 \\
            \cline{2-4}
            & Fake & 23 & 77 \\
        \hline
        FNN 
            & Real & 23 & 77 \\
            \cline{2-4}
            & Fake & 23 & 77 \\
        \hline
        Buzz
            & Real & 27 & 73 \\
            \cline{2-4}
            & Fake & 27 & 73 \\
        \hline
        \end{tabular}

        \caption{Abundance Results (in \%)}
        \label{tab:relabundance}
        \end{table}
\begin{table*}[!htb]
    \begin{minipage}{.4\textwidth}
      \centering
    \begin{tabular}{|c|c|c|c|c|c|}
        \hline
        {\bf Data} & {\bf Ver.} & \multicolumn{2}{c|}{Female} & \multicolumn{2}{c|}{Male} \\
        \cline{3-6}
        & & Emo. & Poli. &Emo. & Poli. \\
        \hline
        ISOT
            & Real & 37.1 & 62.9 & 35.6& 64.4 \\
            \cline{2-6} 
            & Fake & 48.8 & 51.2 & 46.2 & 53.8 \\
        \hline
        
        LIAR
            & Real & 42.4 & 57.6 & 42.2 & 57.8 \\
            \cline{2-6} 
            & Fake & 42.8 & 57.2 & 43.0 & 57.0 \\
        \hline
        FNN
            & Real & 42.2 & 57.8 & 42.5& 57.5 \\
            \cline{2-6} 
            & Fake & 43.1 & 56.9 & 43.1 & 56.9 \\
        \hline
        Buzz
            & Real & 27.3 & 72.7 & 23.8& 76.2 \\
            \cline{2-6} 
            & Fake & 30.4 & 69.6 & 25.4 & 74.6 \\
        \hline
        \end{tabular}

        \caption{Emotion/Political Densities (in \%)}
        \label{tab:emodensity}
    \end{minipage}     
    \begin{minipage}{.46\textwidth}
      \centering
    \begin{tabular}{|c|c|c|c|c|c|c|c|}
        \hline
        {\bf Data} & {\bf Ver.} & \multicolumn{2}{c|}{Positive} & \multicolumn{2}{c|}{Neutral} & \multicolumn{2}{c|}{Negative} \\
        \cline{3-8}
        & & \hspace*{1.5mm}F\hspace*{1.5mm} & M & \hspace*{1.5mm}F\hspace*{1.5mm} & M & \hspace*{1.5mm}F\hspace*{1.5mm} & M \\
        \hline      
        ISOT
            & Real & 35.2 & 36.2 & 35.2 & 36.2 & 29.6 & 27.5 \\
            \cline{2-8}
            & Fake & 33.1 & 33.5 & 33.1 & 33.5 & 33.8 & 33.0 \\
        \hline
        LIAR
            & Real & 36.4 & 37.0 & 36.5 & 37.0 & 27.1 & 26.0 \\
            \cline{2-8}
            & Fake & 33.8 & 35.3 & 33.9 & 35.3 & 32.3 & 29.4 \\
        \hline
        FNN
            & Real & 36.9 & 37.1 & 36.9 & 37.1 & 26.3 & 25.8 \\
            \cline{2-8}
            & Fake & 35.9 & 36.1 & 35.9 & 36.9 & 28.2 & 27.9 \\
        \hline
        Buzz
            & Real & 35.3 & 34.9 & 35.3 & 34.9 & 32.8 & 30.3 \\
            \cline{2-8}
            & Fake & 32.1 & 30.5 & 32.6 & 30.8 & 35.3 & 38.7 \\
        \hline
        
        \end{tabular}
          
        \caption{Sentiment Profiles (in \%)}
        \label{tab:senprofiles}
    \end{minipage}
    \vspace{0.1in}
\end{table*}

\subsection{RQ1: Abundance Analysis}

The relative abundance of genders (in percentage) across fake and real news, computed using a normalised sum-based aggregation of $N_F$ and $N_W$ over fake and real subsets of the datasets, is found to be similar across the various datasets. The results align with contemporary understandings of severe under-representation of women in news media~\cite{shor2019large}. The average value of female representation ($25\pm 2.5\% $) and male representation ($75\pm 2.5\%$) is found to be statistically significant and consistent ($p < 0.05$) in all the four datasets. This points to a serious issue since these datasets are commonly used for constructing ML models for research in this area. The consequences of over-representation of males in the design of an artificial intelligence models in the media domain could quietly undo substantive parts of decades of advances in the gender equality~\cite{10.1145/3195570.3195580}.


\begin{table*}[!htb]
      \centering
        \begin{tabular}{|c|c|c|}
        \hline
         {} & {\bf Female} & {\bf Male} \\
        \hline
        Real &
              governor, spokesman, chairman, statement, adviser, &governor,spokesman,chairman,statement,adviser \\  
              & meeting, senior, 	rival, 	deputy, ambassador,&  meeting,senior,rival,deputy,ambassador\\
            \cline{1-3}
             Fake & wife,interview,attack,march, statement,fact,  & interview,	fact,	march,	photo,governor \\  
             & reported,reality,mother,daughter &	reported,	statement,	morning,	point,	supporters \\
        \hline
        \end{tabular}
        
        \caption{Most Frequent Words}
        \label{tab:topwords}
\end{table*}

\subsection{RQ2: Affect Analysis}

The relative normalised emotional-political densities - measured as percentage of emotional and political words among proximal words - across genders and news veracities are illustrated in Table~\ref{tab:emodensity}. A clear trend in the result is that female are represented more emotionally and less politically compared with corresponding male in both fake news and real news, when abundance around gender mentions is considered as an indicator of nature of representations. This is consistent in across datasets, with the exception of FNN which shows no significant disparities across gender groups. To mention, BuzzFeed contain news predominantly relating with US election 2016, hence the proximal words are found to be more political than other datasets as expected. It is also notable that BuzzFeed has a very high difference in emotional content for women across real and fake news, clearly consistent with the observations that women in politics are not well portrayed (Ref: Section~\ref{sec:relatedwork}). There is also an obvious and overarching trend that fake news is more emotional and less political compared with corresponding real news.


The sentiment analysis results appear in Table~\ref{tab:senprofiles}. Two broad trends are unmistakably visible. First, there is a shift from positive and neutral sentiments towards negative sentiments when one moves eyeballs from the {\it real} news statistics to {\it fake}. Second, this shift is higher in intensity for {\it female} mentions. These provide evidence to assert the prevalence of negativity towards females and reflecting extant qualitative observations of gendered narratives targeting women in political fake news~\cite{stabile2019sex}.

Table~\ref{tab:emoprofiles} illustrates the relative trends across full emotion profiles computed across eight emotion classes. Apart from the expected deterioration in expressions of {\it trust} in fake news, it is also notable that the prevalence of {\it fear} and {\it anger} are much more than prevalence of {\it joy} and {\it surprise}. These are aligned with current studies that negative emotions are beneficial for fake news virality~\cite{corbu2021fake}. It is notable that, even though there are slight variations in the percentage of  emotions, there are no significant differences in trends across genders.



\begin{table*}[!htb]
    \centering
    \begin{tabular}{|c|c|c|c|c|c|c|c|c|c|c|c|c|c|c|c|c|c|}
        \hline
        {\bf Data} & {\bf Ver.} & \multicolumn{2}{c|}{Anger} & \multicolumn{2}{c|}{Anticipation} & \multicolumn{2}{c|}{Disgust} & \multicolumn{2}{c|}{Fear} & \multicolumn{2}{c|}{Joy} & \multicolumn{2}{c|}{Sadness} & \multicolumn{2}{c|}{Surprise} & \multicolumn{2}{c|}{Trust} \\
        \cline{3-18}
        & & \hspace*{.75mm}F\hspace*{.75mm} & M & \hspace*{3mm}F\hspace*{3mm}  &  M & \hspace*{1.5mm}F\hspace*{1.5mm} & M & F & M & F & M & \hspace*{1.5mm}F\hspace*{1.5mm} & M & \hspace*{1.5mm}F\hspace*{1.5mm} & M & F & M \\
        \hline
        ISOT
            & Real & 10 & 9 &16 &14 & 4 &4 &12 &	11 &6 &	5 &	11 & 9 & 4 & 3 & 38 & 44 \\
            \cline{2-18}
            & Fake & 12 &12 &13 &13 & 7 &7 &14 &	14 &9 &	8 &	11 &10 &5 &	5 &	29 &32  \\
        \hline
        LIAR 
            & Real & 7 &6 &	15 & 13&3 &	3 &	8 &	8 & 8 &	7 &	9 &	8 &	4 &	3 &	45 &51\\
            \cline{2-18}
            & Fake &  8 &8 &14 & 13 & 4 &4 &10 &	9 &	7 &	5 &	8 &	8 &	4 &	3 &	45 &51\\ 
        \hline
        FNN
            & Real & 7 &6 &	15 & 13 & 4 &3 &9 &	8 &	7 &	6 &	8 &	7 &	4 &	3 &	47 &53 \\
            \cline{2-18}
            & Fake & 8 &7 &	14 & 13 & 4 &4 &10 &	9 &	7 &	5 &	8 &	7 &	4 &	3 &	45 &50 \\
        \hline
        Buzz
            & Real & 11 &11 &10 & 12 & 6 &6 &14 & 14 &6 &	5 &	8 &	11 &3 &	3  &34 &39 \\
            \cline{2-18}
            & Fake & 16 &10 &11 & 15 & 6 &8 &15 &15 &6 &5 &	8 &	9 &	3 &	4 &	34 &33 \\
        \hline      
    \end{tabular}
    
    \caption{Emotion Profiles (in \%)}
    \label{tab:emoprofiles}
    \vspace{0.1in}
\end{table*}


\subsection{RQ3: Neighboring Lexicons}

Our analysis of lexical references around gender mentions has been much more revealing. Most words around gender mentions were salutations such as {\it adviser}, {\it governor} and {\it spokesman} in real news and any gendered patterns were overshadowed by such words. This is illustrated in Table~\ref{tab:topwords} which shows the $10$ most frequent words along with male and female in both real news and fake news. When it comes to fake news, words around female mentions contained gendered roles such as {\it wife}, {\it mother} and {\it daughter} in sharp contrast to non-gendered roles such as {\it governor} and {\it ambassador} in real news. This trend was seen to deepen in fake news with words such as {\it love} and {\it pretty} coming up additionally among top female proximal references in {\it fake} news, though beyond the top-$10$. A surprising observation is that {\it attack} happens to be the $3^{rd}$ most common word beside female mentions in fake news, indicating a portrayal of victimization of women in fake news. 

This analysis illustrates that sexist stereotypes relating to women - especially those that objectify or victimize them - have an unmistakably higher role in fake news. The lexical references also indicate that fake news is more oriented towards sensationalism (e.g., see words such as attack, photo etc.), something that has been well understood in society.

\section{Conclusions}

We analyzed gender bias using lexicon-based methods over popular fake news datasets to gather quantitative evidence on the relationship between gender and fake news. Our analysis shows that the issue of gender bias - particularly bias against women - is accentuated within fake news on all four aspects of our analysis viz., abundance, affect, political mentions and word references. This indicates, among other consequences, that gender-agnostic processing of fake news could propagate and/or amplify gender biases because if a data is laden with stereotypical concepts of gender, the resulting application of the technology will perpetuate this bias~\cite{10.1145/3195570.3195580}. Our insights provide a strong argument to consider gender bias across the gamut of fake news research. 

While we restricted our analysis to lexicon/dictionary based methods for transparency, our work provides a solid foundation for further studies such as ML-based analyses to uncover details of relationship between gender bias and fake news. In immediate future work, we intend to study ways of extending this analysis to non-binary genders. Usage of parsing-oriented NLP (e.g., dependency parsing) to address the same questions is another interesting direction.



\bibliographystyle{ACM-Reference-Format}
\bibliography{sgnof}


\end{document}